\documentclass[aps,prb,twocolumn,superscriptaddress,showpacs,floatfix]{revtex4-2}
\pdfoutput=1
\usepackage{graphicx,amsmath,amssymb,bm,xcolor}
\usepackage{color,amsfonts,latexsym}
\usepackage{tikz}
\usepackage{amsmath}

\newcommand{\be}{\begin{equation}}
\newcommand{\ee}{\end{equation}}
\newcommand{\ba}{\begin{eqnarray}}
\newcommand{\ea}{\end{eqnarray}}
\newcommand{\baa}{\begin{eqnarray*}}
\newcommand{\eaa}{\end{eqnarray*}}

\def\be{\begin{equation}}
\def\ee{\end{equation}}
\def\bea{\begin{eqnarray}}
\def\eea{\end{eqnarray}}

\def\C60{A$_x$C$_{60}$}

\def\HgCu3{HgCa$_2$Cu$_3$O$_{8+y}$}
\def\HgCu4{HgBa$_2$Ca$_3$Cu$_4$O$_{10+y}$}
\def\TlCu{Tl$_2$Ba$_2$CuO$_{6+\delta}$}
\def\TlCu3{Tl$_2$Ba$_2$Ca$_2$Cu$_3$O$_{10+y}$}
\def\TlCu4{Tl$_2$Ba$_2$Ca$_3$Cu$_4$O$_{12+y}$}

\def\BiCu3{Bi$_2$Sr$_2$Ca$_{2}$Cu$_3$O$_y$}

\def\8LSCO{La$_{1.88}$Sr$_{.12}$CuO$_4$}
\def\110LNSCO{La$_{1.5}$Nd$_{0.4}$Sr$_{0.1}$CuO$_{4}$}
\def\stage4LCO{La$_{2}$CuO$_{4+\delta}$}
\def\Y248{YBa$_2$Cu$_4$O$_8$}

\def\NbSe2{NbSe$_2$}
\def\TaSe2{TaSe$_2$}
\def\TiSe2{TiSe$_2$}

\begin{document}
\title{Pairing Symmetries of Unconventional High Temperature Superconductivity in a Zinc-Blende Structure}

\author{Xilin Feng}
\affiliation{Beijing National Laboratory for Condensed Matter Physics, and Institute of Physics, Chinese Academy of Sciences, Beijing 100190, China}
\affiliation{School of Physical Sciences, University of Chinese Academy of Sciences, Beijing 100190, China}
\author{Qiang Zhang}\email{q.zhang@iphy.ac.cn}
\affiliation{Beijing National Laboratory for Condensed Matter Physics, and Institute of Physics, Chinese Academy of Sciences, Beijing 100190, China}
\author{Zhongyi Zhang}
\affiliation{Beijing National Laboratory for Condensed Matter Physics, and Institute of Physics, Chinese Academy of Sciences, Beijing 100190, China}
\affiliation{School of Physical Sciences, University of Chinese Academy of Sciences, Beijing 100190, China}
\author{Yuhao Gu}
\affiliation{Beijing National Laboratory for Condensed Matter Physics, and Institute of Physics, Chinese Academy of Sciences, Beijing 100190, China}
\author{Kun Jiang}
\affiliation{Beijing National Laboratory for Condensed Matter Physics, and Institute of Physics, Chinese Academy of Sciences, Beijing 100190, China}
\author{Jiangping Hu}\email{jphu@iphy.ac.cn }
 \affiliation{Beijing National Laboratory for Condensed Matter Physics, and Institute of Physics, Chinese Academy of Sciences, Beijing 100190, China}
\affiliation{Kavli Institute of Theoretical Sciences, University of Chinese Academy of Sciences, Beijing 100049, China}
\affiliation{New Cornerstone Science Laboratory, Beijing 100190, China}

\begin{abstract}
We classify the pairing symmetries of three-dimensional superconductivity in the zinc-blende structure which can support an electronic environment to host unconventional high temperature superconductivity, and  calculate the pairing symmetry  in the presence of strong electron-electron correlation by the  slave boson mean-field approach.  We find that  the $d_{2z^2-x^2-y^2}\pm id_{x^2-y^2}$ pairing state,  a three dimensional analogy of the $d\pm id$ pairing in a two dimensional square lattice, is ubiquitously favored near half filling upon hole doping  in both single-orbital and three-orbital models.  However, unlike the two dimensional counterpart, the Bogoliubov quasiparticle spectrum of  the three dimensional  state upholds the full $T_d$ point group symmetry and encompasses point nodes along certain high symmetric lines.
\end{abstract}

\maketitle

\section{\label{s:intro}Introduction}
We recently proposed a new family of unconventional high-temperature superconductors in the zinc-blende structure, guided by the ``gene" concept of high-temperature superconductors\cite{hu2016identifying}. This theory emphasizes that the d-orbitals of the cations involved in strong hybridization and superexchange should be isolated around the Fermi surface. In contrast to the two well-known examples, namely cuprates and iron-based superconductors\cite{lee2006doping,tsuei2000pairing,dai2015antiferromagnetic,dagotto2013colloquium,zaanen2006towards}, the new family of superconductors is primarily three-dimensional in nature. Typically, the orbitals in the 3d-shell of transition metal atoms extend in one or two dimensions. Consequently, it is commonly believed that strong correlation and high-temperature effects occur in quasi-two dimensions.
However, there can be exceptions to this generalization, where specific cation-anion local environments collaborate with the global symmetry. The zinc-blende structure is one such exception.\cite{zhang2020unconventional}. 

The zinc-blende structure possesses a space group of $F\Bar{4}3m (No. 216)$ and a corresponding point group of $T_d$. The crystal field splitting effect results in the upward displacement of the three $t_{2g}$-orbitals, namely $d_{xy}$, $d_{yz}$, and $d_{zx}$. Consequently, these orbitals exhibit strong hopping interactions in three spatial directions. In the vicinity of half filling (with an electronic configuration approximating d$^7$), this system acts as an unconventional three-dimensional high-temperature superconductor. Given the system's high symmetry, the investigation of the superconducting states and their corresponding symmetries in this structure holds great interest.

The investigation of pairing symmetries constrained by lattice symmetries has been a topic of inquiry for a long time \cite{sigrist1991phenomenological}. Specifically, in heavy-fermion superconductivity \cite{assmus1984superconductivity, batlogg1987superconductivity, de1987upt3, varma1985phenomenological}, the $O_h$, $D_{4h}$, and $D_{6h}$ point groups have received significant attention due to their corresponding material-specific properties. However, the $T_d$ group, which features the zinc-blende structure, exhibits different pairing symmetries. In Section \ref{s:sym}, we provide an overview of the representations of the $T_d$ group for convenience in subsequent discussion. In the same Section, we employ the phenomenological Landau theory to explore the full range of possible pairing states determined by the lattice symmetry. To determine the ground state in this model, we use the slave-boson mean field method, as discussed in Section \ref{s:sbmf}. Remarkably, the analysis reveals that the most favorable state of superconductivity is a $d \pm id$ state that breaks time-reversal symmetry but maintains lattice symmetry. This pairing function is a direct three-dimensional analogy to the $d \pm id$ pairing found in quasi-two-dimensional twisted double-layer copper oxides \cite{ZhesenYang, can2021high, volkov2020magic, liu2023high}, but it features zeros along the cubic diagonal lines in the reciprocal lattice, resulting in point nodes in the Bogoliubov quasiparticle spectrum. Lastly, in Section \ref{s:quasi}, we analyze the symmetries of the quasiparticle spectra for various pairing states.

\section{Representations and basis}\label{s:rep}
We present the character table and real-space basis functions $f_i(\mathbf{r})$ for the $T_d$ group in Table \ref{tab:rep} for reference purposes. In our analysis, we focus on the nearest neighbor (NN) hopping and pairing, and as such, the cubic harmonics on the NN bond provide useful insight. Specifically, the zinc-blende structure possesses 12 NN bonds that form lattice harmonics with basis functions $\Gamma_i(\mathbf{k})$ that are classified at the bottom of Table \ref{tab:rep}. Here, $\Gamma$ denotes the representation symbol with values $a_1, e, t_1, \text{ and } t^{o/e}_2$, and the basis functions are given by the expression 
\begin{equation}
\Gamma_i(\mathbf{k})\sim \sum_j e^{i\mathbf{k}\cdot \mathbf{a}_j} f_i(\mathbf{a}_j),    
\end{equation}
where $\mathbf{a}_j$ are the 12 NN lattice vectors. We omit the normalized constant and note that the basis lattice vectors are $\mathbf{a}_1=(1,1,0)/2$, $\mathbf{a}_2=(1,0,1)/2$, and $\mathbf{a}_3=(0,1,1)/2$. Among the representation harmonics, we identify that the basis functions $t_{1i}$ and $t^o_{2i}$ possess spatial parity oddness, and the imaginary factor with the odd parity ensures the time reversal symmetry of the basis functions. The $A_2$ representation cannot be expressed within the NN harmonics. We can obtain further neighbor cubic harmonics similarly. Specifically, the second NN only has the $a_1(\mathbf{k})$ and $e_i(\mathbf{k})$ forms as given in Table \ref{tab:rep}, whereas the third NN harmonics can be obtained by replacing $\frac{k_\mu}{2}$ with $k_\mu$ in NN harmonics.
\begin{table}[t]
    \caption{Character table and basis functions for T$_d$ point group. The NN and second NN cubic harmonics are also listed below  as the basis functions of representations,  where the NN $c_\mu$ and $s_\mu$ represent $\cos{(k_\mu/2)}$ and $\sin{(k_\mu/2)}$, respectively, and the second NN is represented by $C_\mu$ with its value being $\cos{k_\mu}$. The variable $\mu$ takes the values of x, y, and z. The symbol $t^p_{2i}$, where $p$ equals o or e, denotes odd or even parity $t_{2i}$ basis functions.}
    \centering
    \begin{ruledtabular}
    \begin{tabular}{c||c|c|c|c|c|c}
        $T_d$   & E &  8$C_3$   & 3$C_2$    &6$S_4$     & 6$\sigma_d$   & basis functions $f_i(\mathbf{r})$ \\\hline 
        $A_1$   &  1&  1        &   1       &  1        &   1           &    1,$x^2+y^2+z^2$,$xyz$  \\
        $A_2$   & 1 &  1        &   1       &  -1       &  -1           & $ \begin{array}{cc}
             & x^4(y^2-z^2)+y^4(z^2-x^2)  \\
             & +z^4(x^2-y^2)
        \end{array}$ \\
        $E$     & 2 & -1        &   2       &   0       &   0           &$[2z^2-x^2-y^2,\sqrt{3}(x^2-y^2)]$\\
        $T_1$   & 3 &  0        &  -1       &   1       &   -1          &$[x(y^2-z^2),y(z^2-x^2),z(x^2-y^2)]$\\
        $T_2$   &  3&  0        &   -1      &   -1      &   1           &$[x,y,z],[yz,xz,xy]$\\
        \hline\hline
       \multicolumn{7}{c}{ Cubic Harmonics $\Gamma_i(\mathbf{k})$ }\\\hline
        $a_1$    &\multicolumn{6}{c}{$c_xc_y+c_yc_z+c_zc_x$, $C_x+C_y+C_z$}\\
        $e_i$    &\multicolumn{6}{c}{$\begin{array}{cc}
             & [2c_xc_y-c_yc_z-c_xc_z,\sqrt{3}(c_yc_z-c_xc_z)],  \\
             & [2C_z-C_x-C_y,\sqrt{3}(C_x-C_y)]
        \end{array} $}\\
        $t_{1i}$    &\multicolumn{6}{c}{$[is_x(c_y-c_z),is_y(c_z-c_x),is_z(c_x-c_y)]$}\\
        $t^{o}_{2i}$    &\multicolumn{6}{c}{$[is_x(c_y+c_z),is_y(c_z+c_x),is_z(c_x+c_y)]$}\\
        $t^e_{2i}$       & \multicolumn{6}{c}{$[s_ys_z,s_xs_z,s_xs_y]$}
        
    \end{tabular}
    \end{ruledtabular}
    \label{tab:rep}
\end{table}

We also tabulate the product of the $T_d$ group representations in Table \ref{tab:rep_pro} for future reference. In Section \ref{s:sym}, we employ the four-fold products of these representations in the free energy, and even six-fold order terms, to distinguish between potentially degenerate superconducting states. In the three-orbital model, the bilinear forms of the orbital component must be decomposed into different representations. This process involves decomposing the $T_2 \otimes T_2$ representation, as listed in Table \ref{tab:rep_pro}. Additionally, for the pairing representation in the three-orbital model, we consider the product of the cubic harmonics and the bilinear forms of the orbitals, with the appropriate decomposition listed in Table \ref{tab:screp}.

\begin{table}[b]
\caption{Product table for $T_d$ point group}
    \centering
    \begin{ruledtabular}
    \begin{tabular}{c||c|c|c|c}
          &$A_2$  & $E$   &   $T_1$   &   $T_2$     \\\hline
        $A_2$ &$A_1$  & $E$   &   $T_2$   &   $T_1$   \\
        $E$   & $E$  & $A_1+A_2+E$   &   $T_1+T_2$   &   $T_1+T_2$   \\
        $T_1$ & $T_2$  & $T_1+T_2$   &   $A_1+E+T_1+T_2$   &   $A_2+E+T_1+T_2$   \\        
        $T_2$ & $T_1$  & $T_1+T_2$   &   $A_2+E+T_1+T_2$   &   $A_1+E+T_1+T_2$   \\            
    \end{tabular}
    \end{ruledtabular}
    \label{tab:rep_pro}
\end{table}

\begin{figure}
    \centering
    \includegraphics[width=0.49\textwidth]{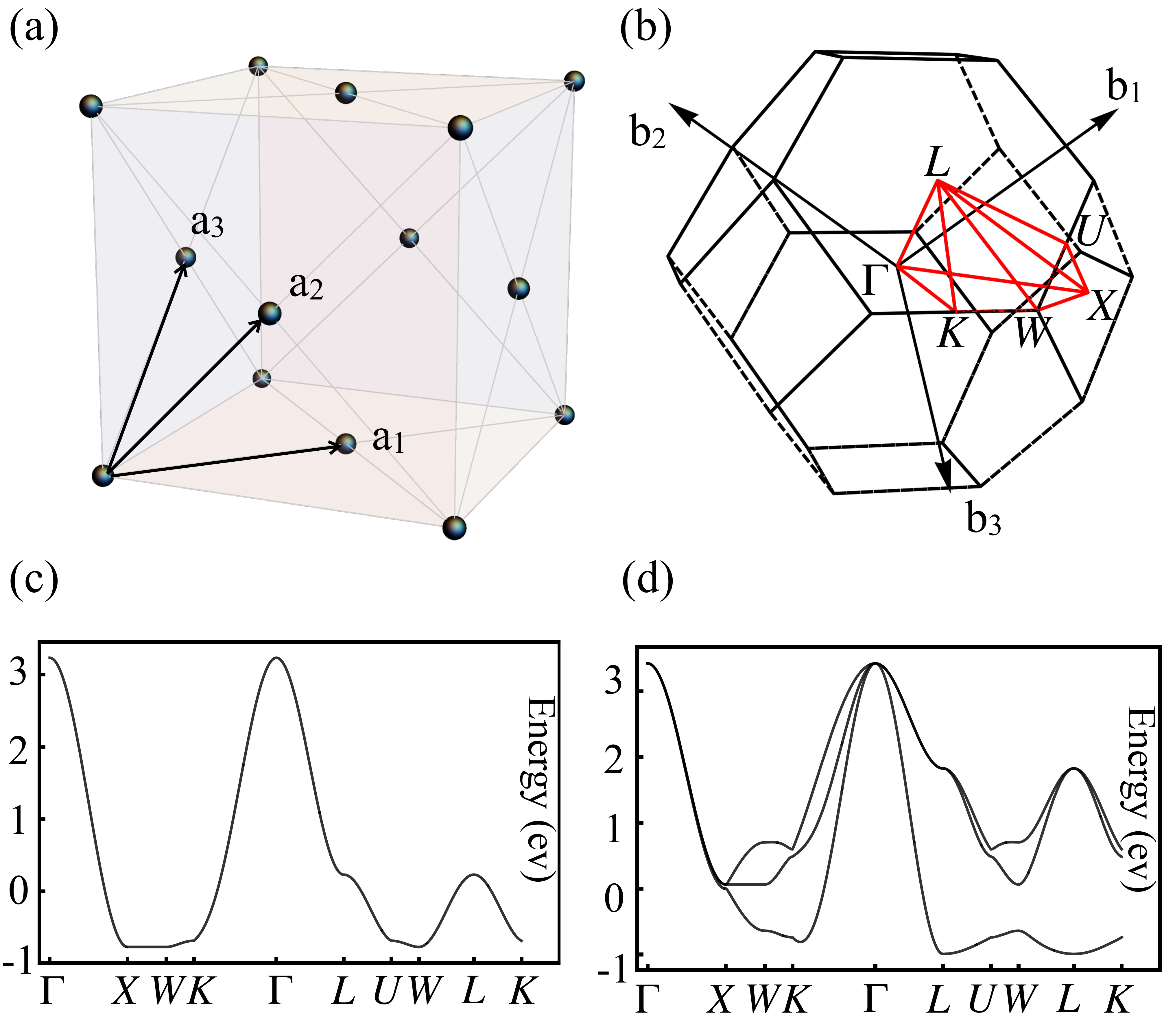}
    \caption{(a). The face center cubic (FCC) structure and the unit vectors in real space. (b). The Brillouin zone of FCC structure and the high-symmetry lines. (c). The dispersion of the single-orbital tight binding model along high-symmetry lines. (d). The dispersion of the three-orbital tight binding model along high-symmetry lines.}
    \label{BZdis}
\end{figure}

\section{Symmetry of Superconductivity }\label{s:sym}

\subsection{Single-orbital model}\label{ssec:equ}
For clarity, we start with a single-orbital model. The cation atoms form a face-centered cubic (FCC) structure, which is illustrated in Fig. \ref{BZdis}(a) alongside its respective unit vectors. Its point group is $O_h$, which is higher than the $T_d$ group for the extra inversion symmetry. Nevertheless, the parity-even representation basis functions and product table coincide with those of $T_d$.  The NN hopping can be described by the Hamiltonian $H=\sum_{k,\sigma}\hat{ a}^\dagger_{k\sigma} H_{1t}\hat{ a}_{k\sigma}$, where $\sigma$ represents spin-up or spin-down, and
\begin{align}\label{equ:1tb}
    H_{1t}=4ta_1(\mathbf{k})-\mu.
\end{align}
Here $t=1/4$ denotes the nearest neighbor hopping amplitude.  At half filling, the chemical potential $\mu$ satisfies $\mu/t = -0.92$. Additionally,  $H_k$ belongs to $A_1$ representation of the point group. The Brillouin zone of FCC structure and the single-orbital tight binding model's dispersion along high-symmetry lines are demonstrated in Fig.\ref{BZdis} (b) and (c), respectively.

The pairing function belongs to a particular representation of the corresponding group. It is expressed as a linear combination of the cubic harmonics listed in Table.\ref{tab:rep}.  In our previous work\cite{zhang2020unconventional}, we only consider the equal pairing amplitude $\delta_{a/e/t_2}$ on the NN bonds which only allows 
\begin{align}
    \Delta^{A_{1}}(\mathbf{k})&=4\delta_a a_1(\mathbf{k}),\nonumber \\
    \Delta^{E_{\pm}}(\mathbf{k})&= 2\delta_e (e_1(\mathbf{k})\pm i e_2(\mathbf{k})), \\
    \Delta^{T_{2}}(\mathbf{k})&= 4\delta_{t_2} (t^e_{21}(\mathbf{k})+e^{i\theta_1}t^e_{22}(\mathbf{k})+e^{i\theta_2}t^e_{23}(\mathbf{k})).\nonumber
\end{align}
It should be noted that the condition of the equal NN bond pairing amplitude excludes the odd parity pairing. It also fixes the amplitude and phase in $E$ representation, leading to the $d_{2z^2-x^2-y^2}\pm i d_{x^2-y^2}$ wave. However, for the $T_2$ representation, there are still two-phase freedoms, represented by $\theta_{1,2}$, which need to be determined using the phenomenological Ginzburg-Landau theory.

\subsection{Landau theory}\label{ssec:landau}
According to Landau's phase transition theory, symmetries are broken at the critical temperature. In conventional superconductivity, only the $U(1)$ symmetry is broken, while unconventional pairing states can spontaneously break additional symmetries such as time reversal, spin rotation, and point group symmetry. The pairing form can be obtained by solving the mean-field self-consistent equation, which involves an eigenvalue problem with the gap function representing the corresponding eigenfunction. Pairings belonging to different representations have different transition temperatures. Therefore, each pairing form carries a specific representation and can be expressed as a linear combination of wavefunctions $\Gamma_i$, corresponding to that representation, with coefficients $\delta_i$.  At high temperature, these coefficients are zero to fulfill the complete symmetry, while at temperature lower than the critical temperature, their non-zero values break some symmetries. 

As a function of the wave-functions, the free energy respects all the symmetries including point group, time reversal, $U(1)$ symmetry and spin rotational $SU(2)$ symmetry. To maintain the $U(1)$ and time reversal symmetries, only real even-order products of $\delta_i$'s are allowed, with the second-order term being $\alpha \sum_{i}|\delta_i|^2$. To obtain the fourth-order terms, we extract terms belonging to $A_1$ representation from $\Gamma \otimes \Gamma^\ast \otimes \Gamma \otimes \Gamma^\ast$. Since this process for $A_1$ representation is trivial, the $E$ and $T_2$ representations are considered below.

\subsubsection{E representation}
In the $E$ representation, the pairing form is given by $\Delta(\mathbf{k})=2\sum_{i=1,2}\delta_i e_i(\mathbf{k})$. Referring to Table \ref{tab:rep_pro} for the product of representations, we can list three terms belonging to $A_{1}$ representation as follows:
\begin{flalign}
&\  \ (|\delta_1|^2+|\delta_2|^2)^2,\nonumber\\
&\  \ (\delta_1\delta^\ast_2-\delta^\ast_1\delta_2)^2,\\
&\  \ (\delta_1\delta^\ast_1-\delta^\ast_2\delta_2)^2+(\delta_1\delta^\ast_2+\delta^\ast_1\delta_2)^2.\nonumber&
\end{flalign}
The third term is not independent, thus the fourth-order term in the free energy is given by:
\begin{align}\label{equ:freeE}
    \Delta F^E_4=\beta_1 (|\delta_1|^2+|\delta_2|^2)^2+\beta_2 (\delta_1\delta^\ast_2-\delta^\ast_1\delta_2)^2.
\end{align}
We can substitute the pairing parameters $(\delta_1,\delta_2)=\delta_0(\cos{\gamma},\sin{\gamma}e^{i\theta})$ to rewrite the fourth-order terms in the free energy, which are given by:
\begin{align}
    \Delta F^E_4=\delta^4_0(\beta_1-\beta_2\sin^2\theta\sin^2(2\gamma)). 
\end{align}

In two different cases, the minimum of the free energy corresponds to two stable phases, which are $e_1(\mathbf{k})$, $e_2(\mathbf{k})$, or $e_1(\mathbf{k})+ce_2(\mathbf{k})$ ($c\in\mathbb{R}$) when $\beta_2<0$, and $e_1(\mathbf{k})\pm i e_2(\mathbf{k})$ when $\beta_2>0$ and $\beta_1-\beta_2>0$. Considering only fourth-order terms in the phenomenological theory results in degeneracy between the $e_1$ state and the $e_2$ state. This degeneracy can be eliminated by introducing six-order terms in the free energy or through the slave-boson mean field calculation. Additionally, in the case of the $T_d$ group, the $e_1/e_2$ basis can cyclically rotate among $x$, $y$, and $z$ due to the $C_3$ rotation, leading to a three-fold degeneracy and $D_{2d}$ subgroup symmetry of the quasiparticle excitation in this state, as shown in Table \ref{tab:subsymmetry1}. The time-reversal-symmetry-breaking (TRSB) $e_1\pm ie_2$ states maintain equal on-bond pairing amplitude, and their Bogoliubov quasiparticle spectra respect the full point group symmetry. This is because the pairing amplitude $\sqrt{e_1^2+e_2^2}$ belongs to $A_1$ representation of the $T_d$ point group.

\subsubsection{T$_2$ representation}
For the $T_2$ representation with $\Delta(\mathbf{k})=4\sum_{i=1-3}\delta_i t^e_{2i}(\mathbf{k})$, we can similarly obtain the terms belonging to $A_1$ representation from the four-representation product $\Gamma\otimes \Gamma^\ast\otimes \Gamma\otimes \Gamma^\ast$. These terms are shown below:
\begin{equation}
    \begin{aligned}
&(|\delta_1|^2+|\delta_2|^2+|\delta_3|^2)^2,\\
&|\delta_1|^4+|\delta_3|^4+|\delta_3|^4-(|\delta_2\delta_3|^2+|\delta_1\delta_3|^2+|\delta_2\delta_1|^2),\\
&(\delta_1\delta^\ast_2\pm \delta^\ast_1\delta_2)^2+(\delta_1\delta^\ast_3\pm \delta^\ast_1\delta_3)^2+(\delta_3\delta^\ast_2\pm \delta^\ast_2\delta_3)^2.
\end{aligned}
\end{equation}

\begin{figure}[t]
     \centering
     \includegraphics[width=0.48\textwidth]{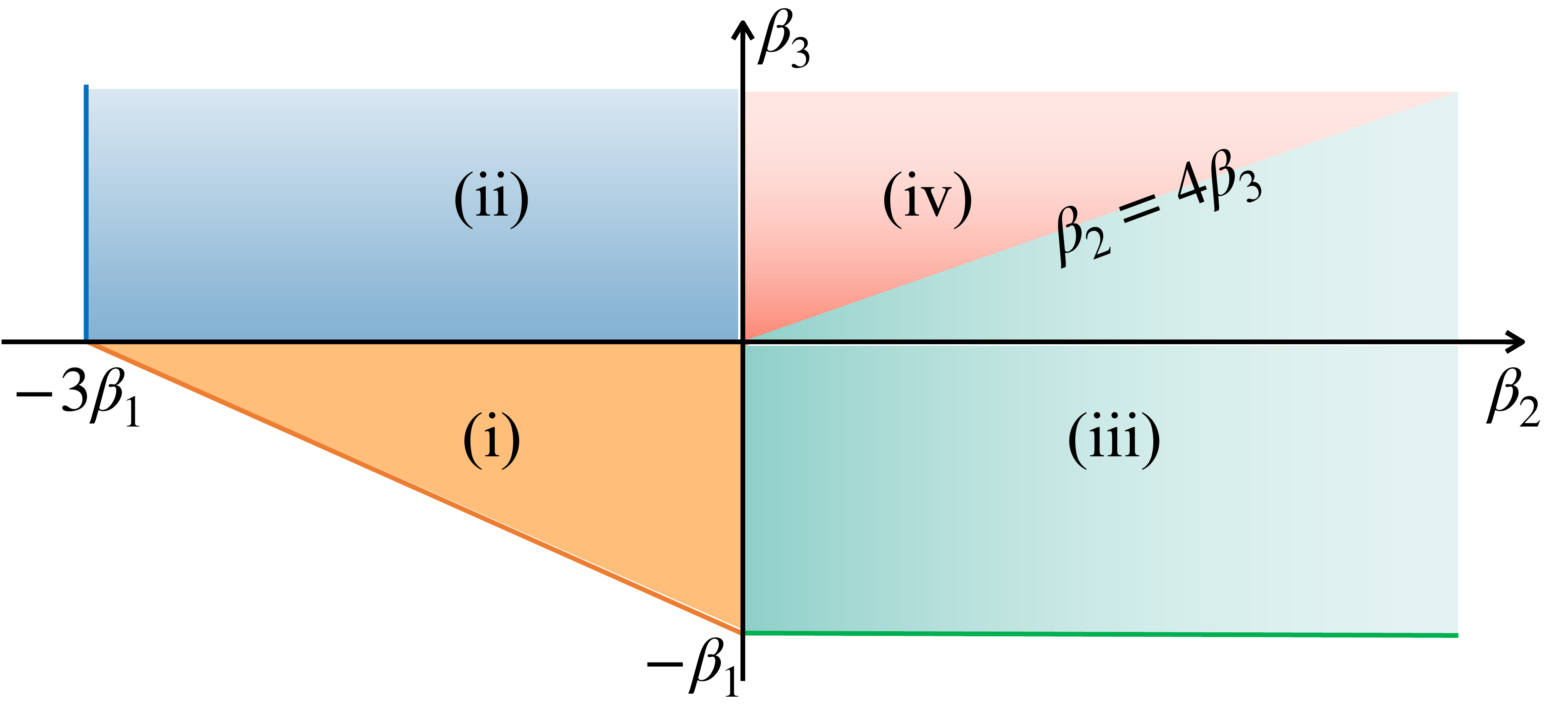}
     \caption{The stable phases from Landau theory for $T_{2}$ representation. The circled numbers denote the time reversal invariant, time reversal broken/chiral, polarized and planar phases, respectively. }
     \label{fig:landau}
 \end{figure}

They are essentially the polynomial invariants that result from projecting $|\delta_1+\delta_2+\delta_3|^4$ onto the basis functions. There are three independent terms. Therefore, the fourth-order terms in free energy can be expressed as follows with some rephrasing:
\begin{align}\label{equ:freeT2}
\Delta F^{T_2}_4& =\beta_1(|\delta_1|^2+|\delta_2|^2+|\delta_3|^2)^2+\beta_2(|\delta_2\delta_3|^2+|\delta_1\delta_3|^2\nonumber\\
&+|\delta_2\delta_1|^2) + \beta_3|\delta^2_1+\delta^2_2+\delta^2_3|^2.
\end{align}
By normalizing $(|\delta_1|^2+|\delta_2|^2+|\delta_3|^2)=1$, the minimum of the free energy yields the four stable  phases    depicted in Fig.\ref{fig:landau}:

\begin{itemize}

     \item (i).  Time reversal invariant $t_{21}\pm t_{22}\pm t_{23}$, when $\beta_3<0$, $\beta_2<0$ and $3\beta_1+\beta_2+3\beta_3>0$

    \item (ii).  Chiral $t_{21}\pm \omega t_{22}\pm \omega^\ast t_{23}$ with $\omega= e^{\pm i2\pi/3}$,  $\beta_3>0$, $\beta_2<0$ and $3\beta_1+\beta_2>0$
    
   \item  (iii). Polarized $(t_{21},t_{22},t_{23})$, when $\beta_3>-\beta_1$, $\beta_2>0$ and $\beta_2-4\beta_3>0$
    
   \item  (iv).  Planar $t_{21}\pm i t_{22}$, when $\beta_2>0$ and $\beta_2-4\beta_3<0$.

\end{itemize}
The degeneracy of each of the four stable phases is 4,8,3,and 6 respectively. Both the time reversal invariant (TRI) and chiral phases have equal on-bond-pairing amplitudes. The point group symmetry is always broken in the quasiparticle spectra of all these pairing forms in the $T_2$ representation.
 
These parameters $\beta_{i}$ can be affected by the spin-fluctuation-feedback effect \cite{amin2020generalized}, which can lead to the transition between different phases in E or $T_{2}$ representations.

\subsection{Three-orbital Model}\label{ssec:3}
\subsubsection{Tight-binding Model}
 \begin{table}[b]
      \caption{Matrix representation basis of the form $ \Gamma_{i,\alpha\beta}$ .}
     \centering
     \begin{ruledtabular}
     \begin{tabular}{c|ccc}
        $\hat{A}_1,\hat{E}_i$  & $\left[\begin{array}{ccc}1&0&0\\ 0&1&0 \\0&0&1\end{array} \right]$ ,   & $\left[\begin{array}{ccc}-1&0&0\\ 0&-1&0 \\0&0&2\end{array} \right]$  & $\left[\begin{array}{ccc}\sqrt{3}&0&0\\ 0&-\sqrt{3}&0 \\0&0&0\end{array} \right]$  \\\hline
        $\hat{T}_{1i}$  & $\left[\begin{array}{ccc}0&0&0\\ 0&0&1 \\0&-1&0\end{array} \right]$    & $\left[\begin{array}{ccc}0&0&-1\\ 0&0&0 \\1&0&0\end{array} \right]$  & $\left[\begin{array}{ccc}0&1&0\\ -1&0&0 \\0&0&0\end{array} \right]$  \\\hline
        $\hat{T}_{2i}$  & $\left[\begin{array}{ccc}0&0&0\\ 0&0&1 \\0&1&0\end{array} \right]$    & $\left[\begin{array}{ccc}0&0&1\\ 0&0&0 \\1&0&0\end{array} \right]$  & $\left[\begin{array}{ccc}0&1&0\\ 1&0&0 \\0&0&0\end{array} \right]$  \\
     \end{tabular}
     \end{ruledtabular}
     \label{tab:Repmatrix}
 \end{table}

We will start by analyzing the symmetry of the three orbitals that form the $T_2$ basis of the point group before considering pairing forms. The Hamiltonian consists of the kinetic energy and mean field pairing terms, which are bilinear forms of the orbital operators. Thus, they form $T_2\bigotimes T_2=A_1\bigoplus E\bigoplus T_1\bigoplus T_2$ representations. We will use the basis $\psi^\dagger=(d^\dagger_{yz},d^\dagger_{xz},d^\dagger_{xy})$ without the spin and momentum indices, as they respond similarly to the real space basis $(\hat{x},\hat{y},\hat{z})$ under the operations belonging to the point group. Therefore, we can express the representations explicitly in terms of the orbital part: $\hat{\Gamma}_i=\sum_{\alpha\beta}\psi^\dagger_\alpha \Gamma_{i,\alpha\beta}\psi_{\beta}$. The matrix forms of the representations are presented in Table \ref{tab:Repmatrix}. The three-orbital tight-binding Hamiltonian, denoted as $\Gamma(\mathbf{k})\bigotimes \hat{\Gamma}$, corresponds to the $A_1$ representation of the point group. Additionally, the Hamiltonian is Hermitian, eliminating the possibility of $t^o_2(\mathbf{k})\hat{T}_2$ and yielding four permissible combinations. Thus the Hamiltonian can be expressed as $H=\sum_{\mathbf{k}} \psi^\dagger(\mathbf{k})H_{3t}(\mathbf{k})\psi(\mathbf{k})$ with the matrix 
\begin{align}\label{equ:3tb}
    H_{3t}(\mathbf{k})& =(h_1a_1-\mu) \hat{A}_1+h_2 \sum_i e_i \hat{E}_i\nonumber\\
                    &+h_3\sum_i t^e_{2i} \hat{T}_{2i}+h_4\sum_i t_{1i}\hat{T}_{1i},
\end{align}
where $h_{1-4}$ are real NN hopping parameters and $\mu$ is the chemical potential. The first two matrix elements are obtained as follow: 
\begin{align}
    H_{3t,11}(\mathbf{k})& =4\big(t_1 (c_xc_y+c_xc_z)+t_2c_yc_z\big)-\mu,\nonumber\\
    H_{3t,12}(\mathbf{k})& =4\big(it_4s_z(c_x-c_y) -t_3s_xs_y)\big).
\end{align}
Other terms can be obtained by a cyclical permutation of $x,y,z$. Here, $4(t_1,t_2,t_3,t_4)=(h_1-2h_2,h_1+4h_2,-h_3,h_4)$ to match the real space NN hopping parameters. For example, in $\mathbf{a}_1$ direction, the hopping matrix $T(\mathbf{a}_1)$ is given by
 \begin{align}
    T(\mathbf{a}_1)=\left(\begin{array}{ccc}
       t_1      &t_3    &t_4    \\
       t_3      &t_1    &t_4    \\
      -t_4      &-t_4   &t_2
    \end{array}\right).
\end{align}
The matrix $T(\mathbf{a}_1)$ is invariant under the mirror on the plane $(\Bar{1}10)$, two-fold rotation along the $z$-axis, and time reversal.

From the tight-binding model, we can analytically determine the band splitting at high-symmetry momenta: $\Gamma$, $X$, $W$, $K$, and $L$. These splittings can aid in fitting the four hopping parameters, $h_i$. At the $\Gamma$ point, the energy is always threefold degenerate at $12h_1$. The band splitting at each momenta is as follows: $\delta_X=-48h_2$, $\delta_W=-24h_2\pm8h_4$, and $\delta_L=-12h_3$. Figure \ref{BZdis} (d) shows the dispersion of the three-orbital tight-binding model along high-symmetry lines with only nearest-neighbor hopping terms. To obtain the hopping parameters and chemical potential listed in Table \ref{para}, we fit the results of the density function theory (DFT) for zinc-blende cobalt nitrogen (CoN) as discussed in Ref. \cite{zhang2020unconventional}. 
\begin{table}[t]
\caption{The parameters of the three-orbital tight binding model obtained by fitting the density function theory's (DFT's) results of the zinc-blende cobalt nitrogen (CoN).}
\begin{ruledtabular}
\begin{tabular}{ccccc}
$h_1$&$h_2$&$h_3$&$h_4$&$\mu$\\ \hline
0.845333&-0.005333&-0.94&0.336&-0.889
\end{tabular}
\label{para}
\end{ruledtabular}
\end{table}

The second NN hopping terms are relatively simple since their harmonics only exhibit $A_1$ and $E$ representations, contributing solely to the intra-orbital hoppings. This can be expressed as:
\begin{align}
    H_{3t,2}=h_5a_{1,2}A_1+h_6 \sum_i e_{i,2} E_i.
\end{align}
The variables are defined as $a_{1,2}=\sum_\mu\cos(k_\mu)$, $e_{1,2}=2\cos(k_z)-\cos{k_x}-\cos{k_y}$, and $e_{2,2}=\cos(k_x)-\cos(k_y)$. The third NN hopping terms are similar to the NN bonds since they share the same symmetries. The only difference is that we should modify the k-dependence of the harmonics, which is obtained by replacing $\frac{k_\mu}{2}$ with $k_\mu$.

\subsubsection{Pairing part}

The method of classifying the pairing component is similar to that of the tight-binding model discussed earlier. The overall representation of pairing consists of cubic harmonics and orbital bilinear forms, which is represented by $\Gamma(\mathbf{k})\otimes \hat{\Gamma}$. In the pairing channels, the operator can be expressed as $\hat{\Gamma}_i=\sum_{\alpha\beta}\psi^\dagger_\alpha \Gamma_{i,\alpha\beta} \psi^\dagger_\beta$, and the matrix form $\Gamma_{i,\alpha\beta}$ is identical to that of the kinetic component because the annihilator and creator transform identically under the point group. Similar to the Hermitian condition of $H_t^\dagger(\mathbf{k})=H_t(\mathbf{k})$ and the time reversal symmetric condition of $H_t^\ast(\mathbf{k})=H_t(-{\mathbf{k}})$ for the kinetic component, the pairing term must satisfy the spin-singlet condition of $\Delta^T(\mathbf{k})=\Delta(-{\mathbf{k}})$ because our focus is primarily on the antiferromagnetic(AFM)-induced spin-singlet pairing. Here, $\Delta(\mathbf{k})$ denotes the pairing matrix in the $\psi^\dagger_\mathbf{k}$ basis. Therefore, only the basis functions with odd parity, such as $t^o_{2i}(\mathbf{k})$ and $t_{1i}(\mathbf{k})$ listed in Table \ref{tab:rep}, can combine with anti-symmetric operators $\hat{T}_{1i}$. 

Table \ref{tab:screp} presents the complete representation decomposition of the nearest-neighbor (NN) singlet pairing. The $A_1$ channel is the same as that of the tight-binding part, where the hopping parameters $h_i$ are replaced by pairing parameters $\delta^{A_1}_i$. The entire $E$ pairing channel comprises six parameters, namely $\delta^E_i$. These include three intra-orbital pairing forms and three inter-orbital forms. The $T_2$ representation pairing channel involves seven free parameters $\delta^{T_2}_i$, including two intra-orbital forms. Notably, due to the combination of orbital bilinear forms and cubic harmonics, the three-orbital model also contains $A_2$ and $T_1$ pairing forms.

In Sec.\ref{ssec:landau}, we listed the stable phases of different representations in the one-orbital model. These phases remain the same in the three-orbital model as we consider the representation under the group theory, except for the additional pairing representations of $A_2$ and $T_1$. The $A_2$ representation is one-dimensional, resulting in a unique stable phase, while the $T_1$ representation shares the stable phases of $T_2$ due to the same phenomenological Landau theory. The stable $e\pm ie$ states are enumerated as $\Delta^{E_{\pm}}(\mathbf{k}) = \sum_j \delta^E_j(\hat{E}_{1,j}\pm i \hat{E}_{2,j})$, with $\hat{E}_{i,j}$ representing the $i$-th basis function of the $j$-th $E$-representation basis in Table \ref{tab:screp}. To provide clarity, we present an example:
\begin{align}
\hat{E}_{1,4}& =2t_{13}\hat{T}_{13}-t_{11} \hat{T}_{11}-t_{12}\hat{T}_{12},\\
\hat{E}_{2,4}& =\sqrt{3}(t_{11}\hat{T}_{11}-t_{12}\hat{T}_{12}).
\end{align}

\begin{table}[b]
\caption{\label{tab:screp} Nearest neighbor (NN) singlet pairing representation basis. The implicit  representation basis $(\Gamma_i(\mathbf{k})\hat{\Gamma}_j)$ can be obtained analogously.}
\begin{ruledtabular}
\begin{tabular}{c |c}
 &   basis\\\hline
$A_1$ &     $a_1\hat{A}_1, e_1\hat{E}_1+e_2\hat{E}_2,\sum_i  t^e_{2i}\hat{T}_{2i},\sum_i t_{1i}\hat{T}_{1i} $  \\\hline 
$A_2$ &     $e_1\hat{E}_2-e_2\hat{E}_1, \sum_i t^o_{2i}\hat{T}_{1i} $ \\\hline
 $E$ &     $\begin{array}{cc} &  (e\hat{A}_1),(a_1\hat{E}),(e_1\hat{E}_1-e_2\hat{E}_2,e_1\hat{E}_2+e_2\hat{E}_1),\\
                     &(2t_{13}\hat{T}_{13}-t_{11} \hat{T}_{11}-t_{12}\hat{T}_{12},\sqrt{3}(t_{11}\hat{T}_{11}-t_{12}\hat{T}_{12})),\\
                    & (t^e_{2}\hat{T}_{2}),(t^o_{2}\hat{T}_{1}))
             \end{array}$\\\hline
$T_1$  &    $\begin{array}{cc}& (-t^e_{21}(\sqrt{3}\hat{E}_1+\hat{E}_2), t^e_{22}(\sqrt{3}\hat{E}_1-\hat{E}_2),2t^e_{23}\hat{E}_2), \\                                                       &(t^e_{22}\hat{T}_{23}-t^e_{23}\hat{T}_{22},t^e_{23}\hat{T}_{21}-t^e_{21}\hat{T}_{23},t^e_{21}\hat{T}_{22}-t^e_{22}\hat{T}_{21}),\\
        &(e\hat{T}_2),  (t_1\hat{T}_1),(t^o_2\hat{T}_1)\end{array} $  \\\hline
$T_2$   &    $\begin{array}{cc} & (t^e_{2i}\hat{A}_1),(t^e_{21}(\hat{E}_1-\sqrt{3}\hat{E}_2), t^e_{22}(\hat{E}_1+\sqrt{3}\hat{E}_2),2t^e_{23}\hat{E}_1),\\
& (t^e_{22}\hat{T}_{23}+t^e_{23}\hat{T}_{22},t^e_{23}\hat{T}_{21}+t^e_{21}\hat{T}_{23},t^e_{21}\hat{T}_{22}+t^e_{22}\hat{T}_{21}),\\&(a_1\hat{T}_{2i}),(e\hat{T}_2),(t_1\hat{T}_1),(t^o_2\hat{T}_1) \end{array} $
\end{tabular}
\end{ruledtabular}
\end{table}
\section{Slave-boson Mean Field Method}\label{s:sbmf}

\subsection{Single-orbital Superconductivity}
\begin{figure}
    \centering
    \includegraphics[width=0.48\textwidth]{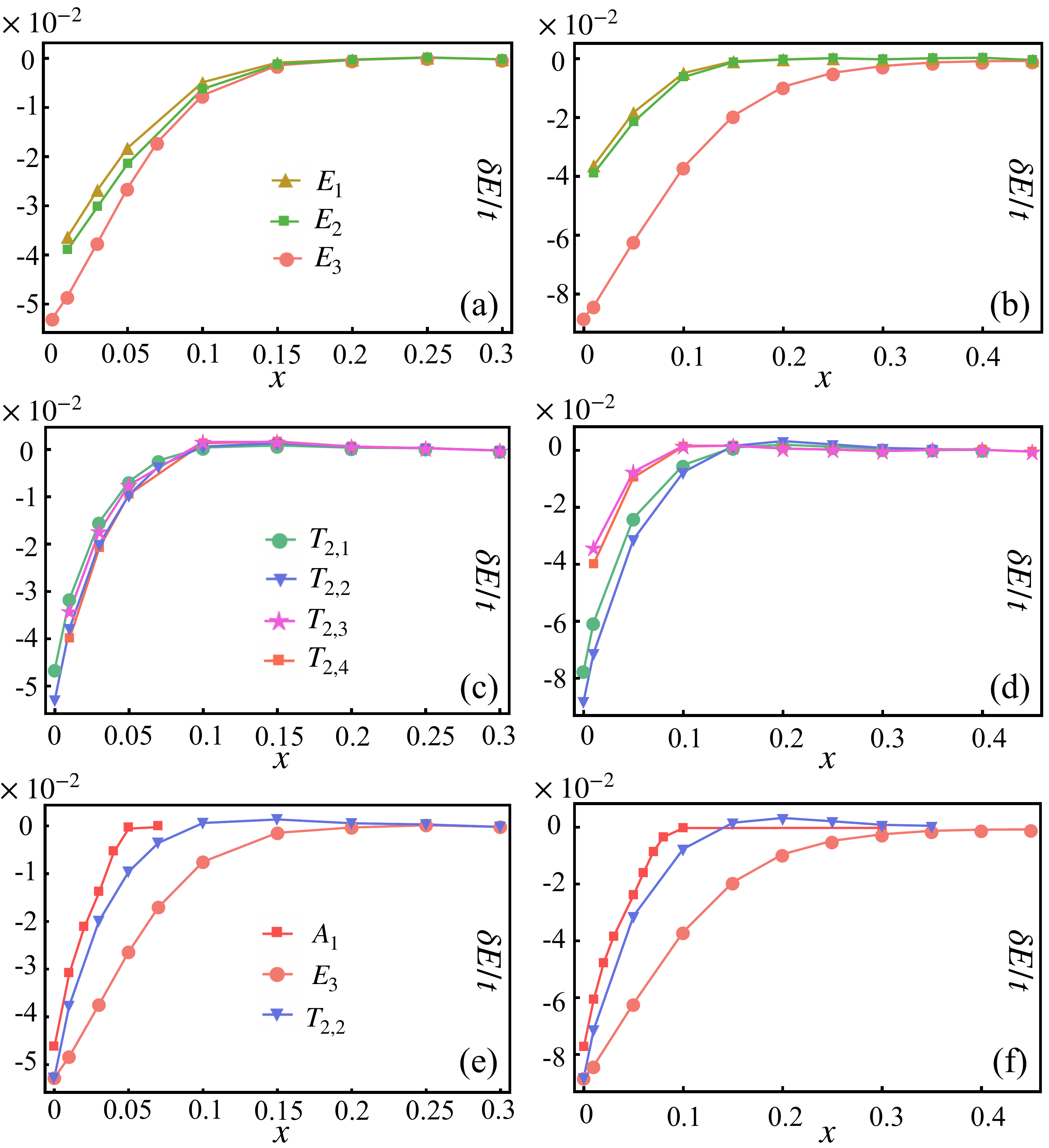}
    \caption{Energy comparison of various superconducting states for single-orbital t-J model around half-filling upon hole-$(x)$ doping. The exchange strength is set  as $J/t=0.3,0.5$ in (a,c,e) and (b,d,f), respectively. The stable states of the $E$ representation are compared in (a,c) and of the $T_2$ representation are in (b,d). }
    \label{fig:sc-1band}
\end{figure}

We employed a single-orbital t-J model \cite{balachandran1990} at half-filling and doped it to investigate the phases of superconductivity, as illustrated in Fig.\ref{fig:sc-1band}. The NN spin-singlet pairing of the one-orbital model only allows $A_1$, $E$, and $T_2$ forms as the spatial component must be even. We set the AFM exchange interaction to $J/t=0.3, 0.5$ in Fig.\ref{fig:sc-1band}. (a,c,e) and (b,d,f), respectively. The energy savings from different pairing terms belonging to the $E$ representation are listed in (a,b), and the results of the terms belonging to the $T_2$ representation are displayed in (c,d). We compared the lowest energy states from different representations in (e,f). Specifically, for the $E$ representation, the pairing form factors of $E_1$, $E_2$, and $E_3$ are denoted as $e_1$, $e_2$, and $e_1 \pm i e_2$, respectively. As for the $T_2$ representations, the pairing form factors of $T_{2,1}$, $T_{2,2}$, $T_{2,3}$, and $T_{2,4}$ are $t_{21}^{e} \pm t_{22}^{e} \pm t_{23}^{e}$, $t_{21}^{e}\pm \omega t_{22}^{e} \pm \omega^2 t_{23}^{e}$, one of $(t_{21}^{e},t_{22}^{e},t_{23}^{e})$, and $t_{2a}^{e}+i t_{2b}^{e}$, where $\omega=e^{\pm i 2\pi/3}$, and $a\neq b$ belong to $(1,2,3)$, respectively.

The multi-component superconductivity generally exhibits lower ground state energy in the time reversal symmetry breaking states. In the $E$ representation, the $d \pm id$ wave has the lowest ground state energy. The states from the $T_2$ representation have nearly indistinguishable energy at a weak exchange interaction of $J/t=0.3$, as shown in (c). However, with stronger antiferromagnetic (AFM) exchange interaction in (d), the chiral states $T_{2,2}$ have the lowest ground state energy. Notably, the $d \pm id$ wave from the $E$ representation maintains the lowest ground state energy upon hole doping, as indicated in (e,f). 

\subsection{Three-orbital Superconductivity}
\begin{figure}[b]
    \centering
    \includegraphics[width=0.48\textwidth]{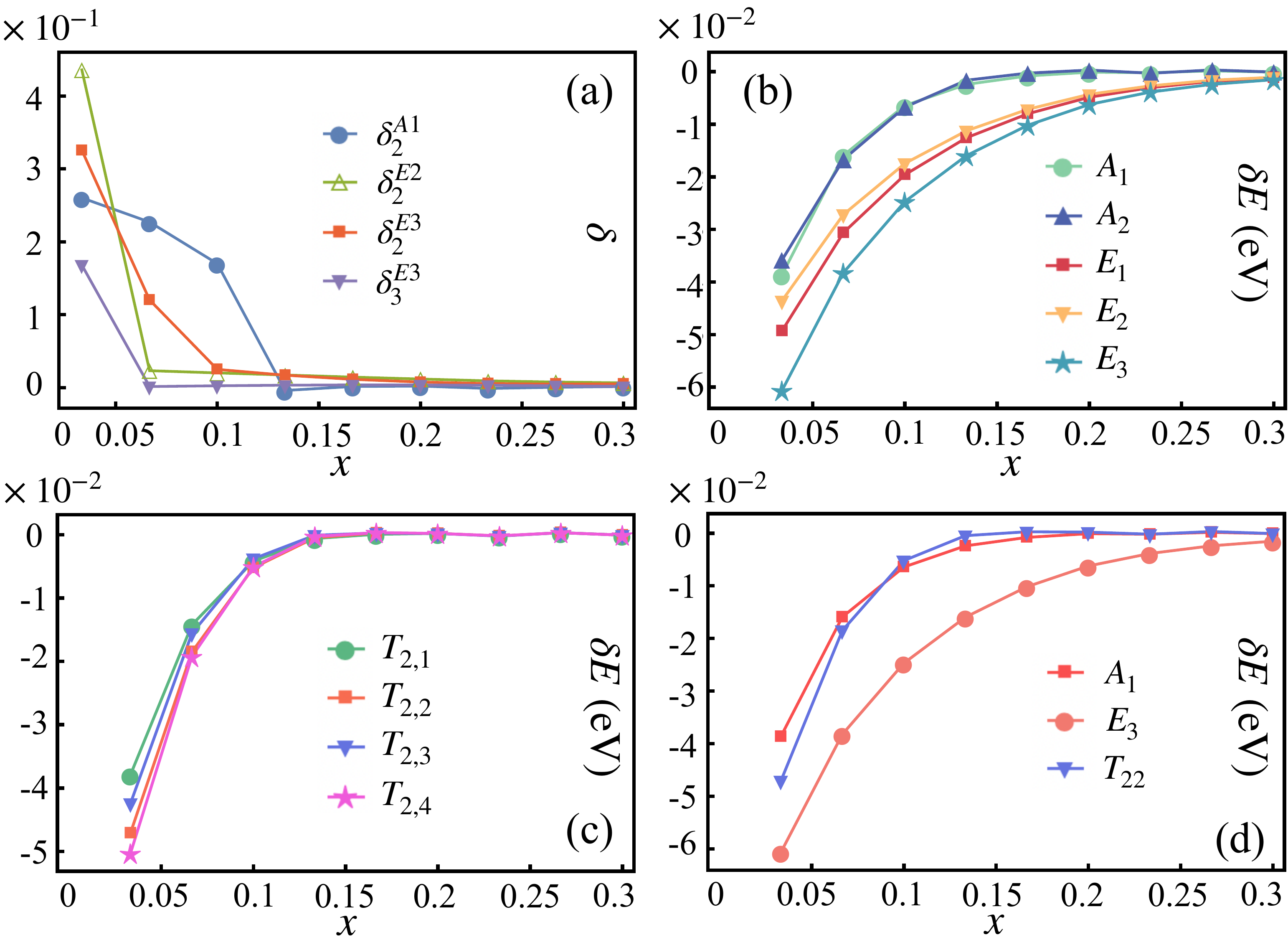}
    \caption{The results for three-orbital t-J model around half filling upon hole doping-$x$. (a) The strength of the anisotropic pairing parameters from different representations. (b,c,d) The energy savings for different superconducting states. The AFM exchange $J=0.1$ eV are adopted.  We use the same notation as with the single orbital case to designate the different pairing states that belong to each representation, since the isotropic intra-orbital pairing channels are dominate, and the matrix forms of their operator parts are always $\hat{A}_{1}$.}
    \label{fig:sc-3band}
\end{figure}

In the three-orbital model, there are multiple pairing parameters in different channels. However, since the inter-orbital pairs generally require more energy to condense, we consider only the pairing with the matrix forms of $\hat{A}_1$ and $\hat{E}$, which refer to intra-orbital channels. In particular, $\hat{A}_1$ signifies the isotropic pairing in the orbital space, which dominates the pairing state. Figure \ref{fig:sc-3band}(a) illustrates the pairing amplitudes $\delta_{2}^{A{1}}$, $\delta_{2}^{E{2}}$, $\delta_{2}^{E{3}}$, and $\delta_{3}^{E{3}}$, which correspond to $e_1\hat{E}_1+e_2\hat{E}_2$, $(a_1\hat{E})$, $(e_1\hat{E}_1-e_2\hat{E}_2,e_1\hat{E}_2+e_2\hat{E}_1)$, and $(t^e_{21}(\hat{E}_1-\sqrt{3}\hat{E}_2), t^e_{22}(\hat{E}_1+\sqrt{3}\hat{E}_2),2t^e_{23}\hat{E}_1)$, respectively. The pairing amplitudes of these anisotropic intra-orbital pairing channels from different representations attenuate rapidly around half-filling upon hole doping. Thus, for the $A_1$, $E$, and $T_2$ representations, we focus only on the isotropic intra-orbital pairing channels. The notations of the pairing channels used in single orbital (Table \ref{tab:subsymmetry1}) can likewise apply to isotropic intra-orbital counterparts in the three-orbital case, since the operator parts of the isotropic intra-orbital pairing channels are always $\hat{A_{1}}$.

The AFM exchange interaction was set at $J=0.1$ eV, approximately half of the hopping strength and 1/24 of the bandwidth. In Figure \ref{fig:sc-3band} (b), we compared the ground state energy of the $A_1$, $A_2$, and the three $E$ states. The two one-dimensional representations possess ground state energies that are nearly degenerate, but neither of them is preferred. The previously degenerate ($E_1$,$E_2$) states are now separated, with the $E_2$ state possessing lower ground state energy. In Figure \ref{fig:sc-3band} (c), the four states from the $T_2$ representation listed in Table \ref{tab:subsymmetry1} are illustrated. They all have almost indistinguishable energies, with slightly lower energies in the TRSB states (chiral and planar). In (d), we compared the lowest energy states in each representation channel. The results demonstrate that the $E_3$ ($d \pm id$) state is the most energetically favorable.

\begin{figure}[b]
    \centering
    \includegraphics[width=0.48\textwidth]{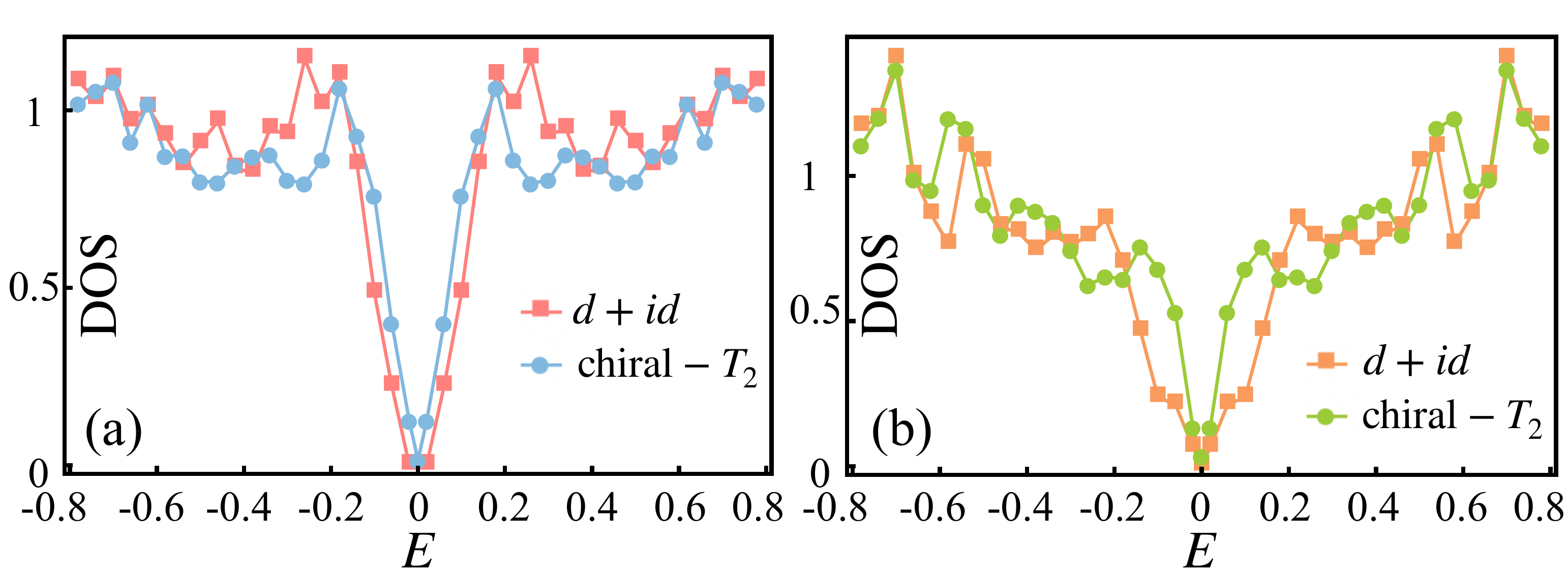}
    \caption{The Density of states (DOS) of the $d\pm id$ and chiral-$T_2$ superconductivity for (a). single-orbital case and (b). three-orbital case, respectively.}
    \label{Dos}
\end{figure}
\begin{table}[t]
\caption{Spectra symmetry of single-orbital pairing channels. The column with identifier $i$ represents the state label. "Deg." abbreviates degeneracy levels, and "Subg" denotes the subgroup symmetry of quasiparticle excitation. The designation "Zeros" indicates the pairing form's zeros. "F" and "L" refer to a two-dimensional face and a one-dimensional line, respectively. The abbreviation "TR" signifies the time reversal symmetry condition. "B" denotes the overall bulk, whereas "F" refers to a specific face.}
\centering
\begin{ruledtabular}
\begin{tabular}{c|c|c|c|c|c}
i & $\Delta(\mathbf{k})$    & Deg.  & Subg. & Zeros & TR\\\hline
$E_1$&  $e_1$   & 3 & $D_{2d}$  & F & B\\
$E_2$&  $e_2$   & 3 & $D_{2d}$  & F & B\\
$E_3$&    $e_1\pm i e_2$ & 2 & $T_d$ & L & F \\
$E_4$&  $e_1+c e_2$ & 3 &   $D_{2(d)}$ & L & B/F\\
\hline
$T_{2,1}$&   $t_{21}^{e}\pm t_{22}^{e}\pm t_{23}^{e}$ & 4 & $C_{3v}$    & L & B\\
$T_{2,2}$&   $t_{21}^{e}\pm \omega t_{22}^{e}\pm \omega^\ast t_{23}^{e}$ & 8 & $C_{3v}$ & L & F\\
$T_{2,3}$&   $(t_{21}^{e},t_{22}^{e},t_{23}^{e})$ & 3 & $D_{2d} $ & F & B\\
$T_{2,4}$&   $t_{2a}^{e}\pm i t_{2b}^{e} (a\neq b) $ & 6 & $D_{2d} $ & F & F\\  
\end{tabular}
\end{ruledtabular}
\label{tab:subsymmetry1}
\end{table}

\section{  quasiparticle excitation}\label{s:quasi}
For the single-orbital pairing channel, the quasiparticle spectra are simply $\pm \sqrt{H_{1t}^2(\mathbf{k})+|\Delta(\mathbf{k})|^2}$, where $H_{1t}(\mathbf{k})$ is the single-orbital normal state dispersion function that is proportional to $A_1$ representation, and invariant under all point group transformations. Therefore, the symmetry of $|\Delta(\mathbf{k})|^2$ governs the quasiparticle spectra, as listed in Table \ref{tab:subsymmetry1}. The enumeration includes the states in each representation and their degeneracy, the symmetry of quasiparticle excitations, zeros of the pairing forms, and the manifolds of time reversal. It should be noted that for $E$ representation, the $O(2)$ symmetry of $\Delta F^E_4(\bm{\delta})$ in Eq.(\ref{equ:freeE}) allows any continuous rotation of the basis $(e_1,e_2)$, rendering the combination of $e_1\pm c e_2$ with any real constant $c$ to be stable according to Landau's phenomenological theory. In addition, when $c$ changes from $0$ to $\sqrt{3}$, the subgroup describing the symmetry of the quasiparticle transforms from $D_{2d}(z)$ to $D_{2d}(x)$. For arbitrary real value of $c$, the subgroup symmetry is $D_{2(d)}$, and it is time reversal invariant throughout the bulk. Regarding the $d\pm id$ ($E_3$) and chiral-$T_2$ ($T_{2,2}$) states, they both have lower ground state energy near half-filling while possessing zero lines and time reversal invariant planes. The zero lines of the $d\pm id$ state are the four bulk diagonal lines of a cube: $|k_x|=|k_y|=|k_z|$. The intersections of these zero lines with the normal state Fermi surface lead to the formation of point nodes on the quasiparticle spectra. The pairing function breaks time reversal symmetry, but preserves time reversal invariant in the planes given by $k_\mu=\frac{\pi}{2}+n\pi$ or $|k_\mu|=|k_\nu|$, with $\mu,\nu=x,y,z$, and $n$ being an integer. As for the chiral-$T_2$ state, the zeroes align on $k_x=k_y=k_z$ and $k_\mu=k_\nu=0$. Similarly, the function preserves time reversal in the planes given by $k_\mu = n\pi$ or $k_\mu=-k_\nu$.  For single-orbital case, the density of states (DOS) of the $d\pm id$ ($E_3$) and chiral-$T_2$ ($T_{2,2}$) superconductivity are shown in Fig.\ref{Dos} (a), which is consistent with the point nodes behavior.
%Particularly, those planes are the time reversal invariant planes, so the recovery of the time reversal symmetry would be of great interest for the study of its possible topological phase in the future.
\begin{figure}[t]
    \centering
    \includegraphics[width=0.5\textwidth]{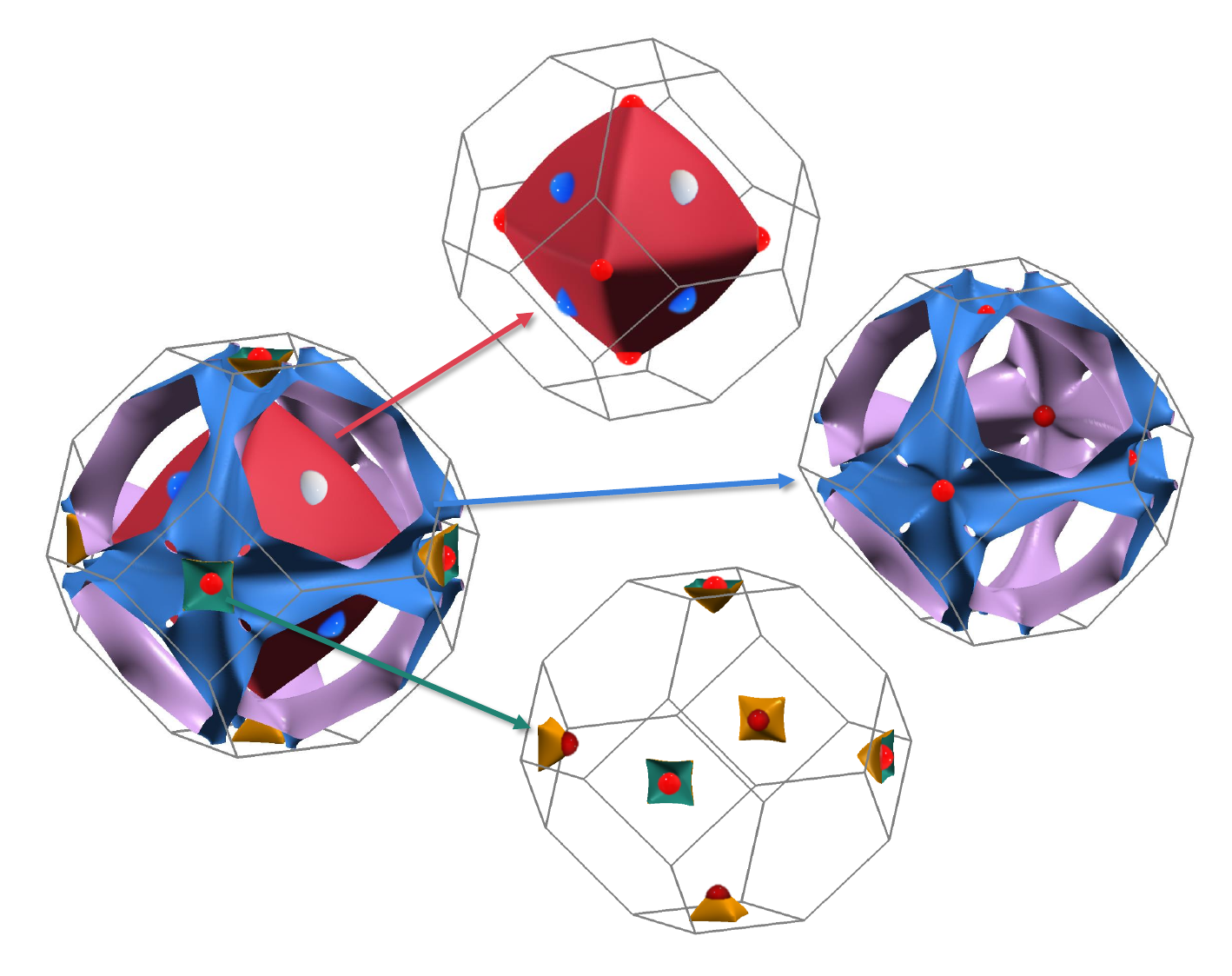}
    \caption{The Fermi surfaces of the three orbital model are obtained by the density function theory (DFT). The point nodes of the $d \pm id$ channel are shown as the blue and white points, while the point nodes of the chiral-$T_{2}$ channel are shown as the red and white points.}
    \label{FS}
\end{figure}

For the three-orbital case, due to the complexity induced by interorbital pairing or anisotropic intra-orbital pairing, the symmetries of the quasiparticle spectra can be different with $|\Delta(\vec{k})|^2$. Here we are also interested in the $d \pm id$  and chiral-$T_2$ states, since they are favored according to the mean-field calculations in Sec.\ref{s:sbmf}. The intersections of the zero lines and the three orbital normal state Fermi surfaces lead to point nodes on quasiparticle spectra, which are shown in Fig.\ref{FS}. Their density of states are shown in Fig.\ref{Dos} (b). The symmetry of quasiparticle spectra is listed in Table.\ref{tab:subsymmetry2}. In this table, for multi-orbital $d \pm id$ and chiral-$T_2$ superconductivity, we also consider the inter-orbital pairing channels and anisotropic intra-orbital pairing channels, and we find that they break more crystal symmetry than isotropic intra-orbital channels. The symmetries of the quasiparticle spectra of the isotropic intra-orbital pairing channels are the same as $|\Delta(\vec{k})|^2$, which is consistent with that in the single-orbital case.

\begin{table}[t]
\caption{The symmetry of the spectra of different representations in three-orbital pairing channel.Only the time-reversal-symmetry-breaking (TRSB) pairing channels that belong to the $E$ and $T_{2}$ representations are considered. The meaning of the notation used for each type of pairing representation basis aligns with those provided in Table.\ref{tab:screp}.}
\centering
\begin{ruledtabular}
\begin{tabular}{c|c|c}
Pairing & With representations	& Subgroup   \\\hline
  &	$(e\hat{A}_{1})$    & $T_{d}$  \\
  &	$(a_{1}\hat{E})$ & $T$   \\
 $d \pm id$ &   $(e_1\hat{E}_1-e_2\hat{E}_2,e_1\hat{E}_2+e_2\hat{E}_1)$ & $T$   \\
  & 	 $\begin{array}{cc} & (2t_{13}\hat{T}_{13}-t_{11} \hat{T}_{11}-t_{12}\hat{T}_{12},\\ &
  \sqrt{3}(t_{11}\hat{T}_{11}-t_{12}\hat{T}_{12}))
         \end{array}$ & $T$  \\
  &   $(t^e_{2}\hat{T}_{2})$  &   $T$ \\
  &   $(t^o_{2}\hat{T}_{1})$  &    $T$\\
\hline
    &   $(t_{2}^{e}\hat{A}_{1})$ & $C_{3v}$ \\
 chiral $T_2$   &   $({t_{2}^{e}\hat{E}}),(e\hat{T}_{2}) $& $C_{3}$ \\
    &  $(a_1\hat{T}_{2i}), (e\hat{T}_2), (t_1\hat{T}_1), (t^o_2\hat{T}_1)$ & $C_{3}$ \\
\end{tabular}
\end{ruledtabular}
\label{tab:subsymmetry2}
\end{table}

\section{Conclusion}\label{s:end}
In summary, we classified the spin-singlet pairing superconductivity in the zinc-blende structure. The $d_{2z^2-x^2-y^2} \pm i d_{x^2-y^2}$ pairing state, as the three-dimensional analogy of the $d \pm id$ pairing in a two dimensional square lattice, was found to be the most stable superconducting phase near half-filling upon hole-doping for both single- and three-orbital models through the slave-boson mean field method. Nonetheless, $d \pm id$ pairings in three dimensions possess line zeroes along the four bulk diagonal lines of a cube, leading to point nodes in quasiparticle spectra. We analyzed the symmetries of quasiparticle spectra belonging to various representations. We found  that the quasiparticle spectra of the TRSB $d \pm i d (E_3)$ pairing in the single-orbital case and the isotropic intra-orbital $d \pm id$ $(e\hat{A}_{1})$ pairing in the three-orbital case maintain the full point-group symmetry. The signal of the time-reversal symmetry breaking in this three-dimensional superconductivity can be detected by muon spin relaxation ($\mu$SR) measurements\cite{amato1997}. The point nodes can  impact the behavior of the specific heat below the superconducting transition temperature\cite{yang2019low} and the temperature-dependence of the superfluid density\cite{hayashi2006temperature}, both of which can be observed in experiments.

{\it Acknowledgement:}  Q.Zhang contributed to this work when he was in IOP. This work is supported by the Ministry
of Science and Technology (Grant No. 2022YFA1403901,
No.2022YFA1403800), the National Natural Science Foundation of China (Grant No. NSFC-11888101, No. NSFC12174428), the Strategic Priority Research Program of the
Chinese Academy of Sciences (Grant No. XDB28000000,
XDB33000000), the New Cornerstone Science Foundation,
and the Chinese Academy of Sciences through the Youth Innovation Promotion Association (Grant No. 2022YSBR-048).

\appendix

\bibliography{ref.bib}
\end{document}